# Solid immersion microlens arrays-based light-field camera for 3D *in vivo* imaging

Jae-Myeong Kwon,[1,2] Sang-In Bae,[1,2] Taehan Kim,[3] Jeong Kun Kim,[3] and Ki-Hun Jeong[1,2,*]

[1]*Department of Bio and Brain Engineering, Korea Advanced Institute of Science and Technology (KAIST), 291 Daehak-ro, Yuseong-gu, Daejeon, 34141, Republic of Korea*
[2]*KAIST Institute for Health Science and Technology (KIHST), KAIST, 291 Daehak-ro, Yuseong-gu, Daejeon, 34141, Republic of Korea*
[3]*Vatech Co. Ltd., 13 Samsung-ro 2-gil, Hwaseong-si, Gyeonggi-do, 18449, Republic of Korea*
*\*kjeong@kaist.ac.kr*

**Abstract:** Light-field imaging facilitates the miniaturization of 3D cameras while it requires the extension of the depth-of-field (DoF) for practical applications such as endoscopy and intraoral scanning. Here we report a light-field camera (LFC) using solid immersion microlens arrays (siMLAs) for 3D biomedical imaging. The experimental results show that the focal length of MLAs is increased by 2.7 times and the transmittance is enhanced up to 6.9% by immersion in PDMS film. In particular, the *f*-number of siMLAs exceeds the limit of conventional MLAs fabricated by thermal reflow, resulting in a larger DoF. The LFC based on siMLAs has successfully acquired the depth map of a dental phantom as a hand-held scanner. This LFC suggests a new way for developing a compact *in vivo* 3D imaging system.

## 1. Introduction

3D imaging techniques have provided essential applications in the medical field. In particular, 3D optical imaging using a stereo camera, structured illumination, or confocal scanning is used for endoscopy and intraoral scanning because depth information enhances diagnostic accuracy, treatment time, and patient comfort [1-3]. However, 3D imaging methods often suffer from large system sizes [1] or motion artifacts [4] since many methods require an additional camera, projector, or moving optical elements. Miniaturization of 3D imaging systems is, therefore, still required for *in vivo* imaging. Light-field cameras (LFCs) facilitate the 3D reconstruction of a scene by obtaining both spatial and angular information of incoming light with a single exposure [5]. The simple configuration of LFCs, consisting of the main lens, microlens arrays (MLAs), and an image sensor, enables easy miniaturization [6] and low motion artifact.

The depth-of-field (DoF) of LFCs is required to be extended for practical 3D imaging [7]. Recently, multi-focal MLAs [8,9] and tunable MLAs [10] are reported to extend the DoF of LFCs while these methods still remain technical issues such as the limited *f*-number range and the restricted DoF in a single exposure. MLAs are often fabricated by thermal reflow of photoresist [8, 11], MEMS-based molding [12], microdroplet [13,14], ultraprecision machining (UPM) [15], or laser manufacturing [16]. Thermal reflow enables the wafer-level fabrication of MLAs [17] and the addition of fabrication steps for more functions such as anti-reflection and [18] optical crosstalk blocking [19,20]. However, the *f*-number of thermally reflowed MLAs hardly exceeds *f*/2.5 [21] while a large *f*-number is required to achieve a large DoF light-field imaging system [22]. This limits applications of LFCs because the *f*-number of a main lens and MLAs should be matched when constructing a light-field imaging system to fully utilize an image sensor [9,23].

In this work, we report solid immersion microlens arrays (siMLAs) for 3D biomedical light-field imaging with a large DoF. The siMLAs are formed by polydimethylsiloxane (PDMS) spin coating after conventional MLAs fabrication by thermal reflow. Solid immersion of MLAs efficiently increases the focal length of MLAs by inserting a lower refractive index medium between MLAs and the air as illustrated in Fig. 1a. Therefore, the siMLAs involving thermal reflow can exceed the *f*-number limit of conventional thermally reflowed MLAs, and this facilitates the *f*-number matching with a larger *f*-number (Fig. 1b) resulting in a larger DoF (Fig. 1c). In addition, the transmittance of a photoresist composing microlenses is increased after PDMS spin coating by reducing reflection at the surface. The siMLAs are microfabricated with high uniformity and mass productivity due to the wafer-level fabrication, and an optical absorber is fabricated to enhance light-field imaging by taking an advantage of thermal reflow. The large *f*-number of siMLAs enables 3D imaging of an actual size dental phantom based on an LFC during hand-held operation, providing a new guideline for practical 3D *in vivo* imaging.

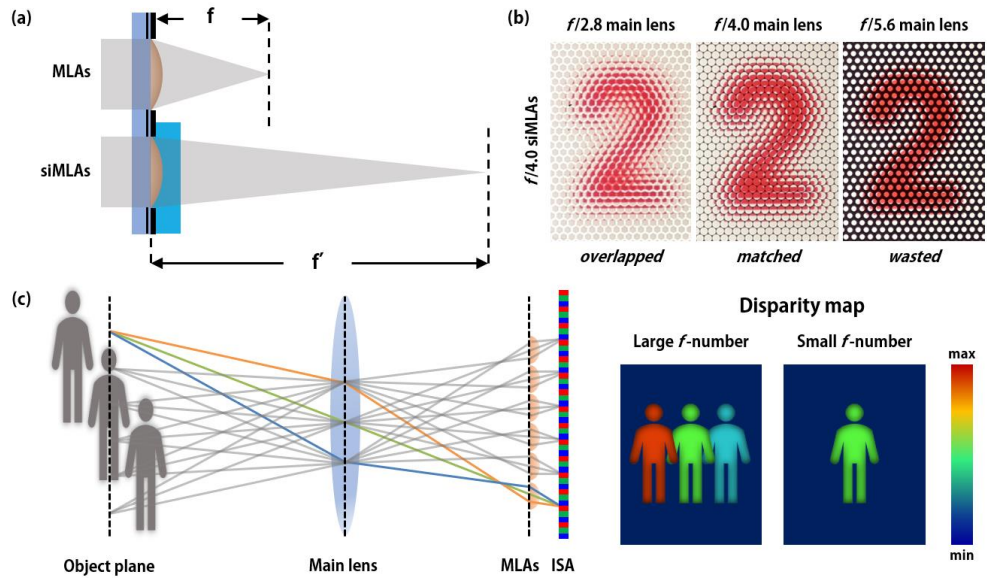

Fig. 1. The siMLAs for light-field imaging. (a) A schematic illustration of siMLAs for the *f*-number extension. Solid immersion of MLAs in PDMS increases the focal length by several times. (b) The *f*-number matching in light-field imaging. The *f*-number of a main lens and MLAs must be matched to fully utilize the image sensor without overlap. (c) A schematic illustration of a light-field imaging system. A large *f*-number is required to achieve a large depth-of-field (DoF) for practical 3D biomedical imaging.

## 2. Result and discussion

### 2.1 Microfabrication of siMLAs

The microfabrication steps of siMLAs are shown in Fig. 2a. A thermoplastic negative photoresist (DNR-L300-D40, Dongjin Semichem Co., Ltd.) is used for repeated lift-off processes and microlens formation. First, the DNR is photolithographically defined on a four-inch borosilicate wafer. The Cr layer is deposited by e-beam evaporation, and the residual DNR is stripped to remove Cr on the place for microlenses. The $SiO_2$ deposition is followed by plasma-enhanced chemical vapor deposition (PECVD), and the DNR deposition and the Cr lift-off are repeated to fabricate a light-absorbing layer blocking optical crosstalk between microlenses. The DNR is defined by photolithography one more time, and MLAs are formed by thermal reflow of the DNR pattern. Finally, MLAs are immersed in PDMS (Sylgard 184,

Dow Inc.; 10:1 base-to-crosslinker ratio) by spin coating, and the PDMS film is cured at 120°C for 20 minutes on a hot plate. The scanning electron microscopic image of microfabricated siMLAs in Fig. 2b shows immersion of MLAs in the PDMS film. Note that some part of the PDMS film is intentionally removed to make a clear distinction between MLAs and immersion material. Microfabricated siMLAs are packaged on a CMOS image sensor to build a light-field imaging system so that the curved surface of microlenses faces the sensor using spacers as illustrated in Fig. 2c.

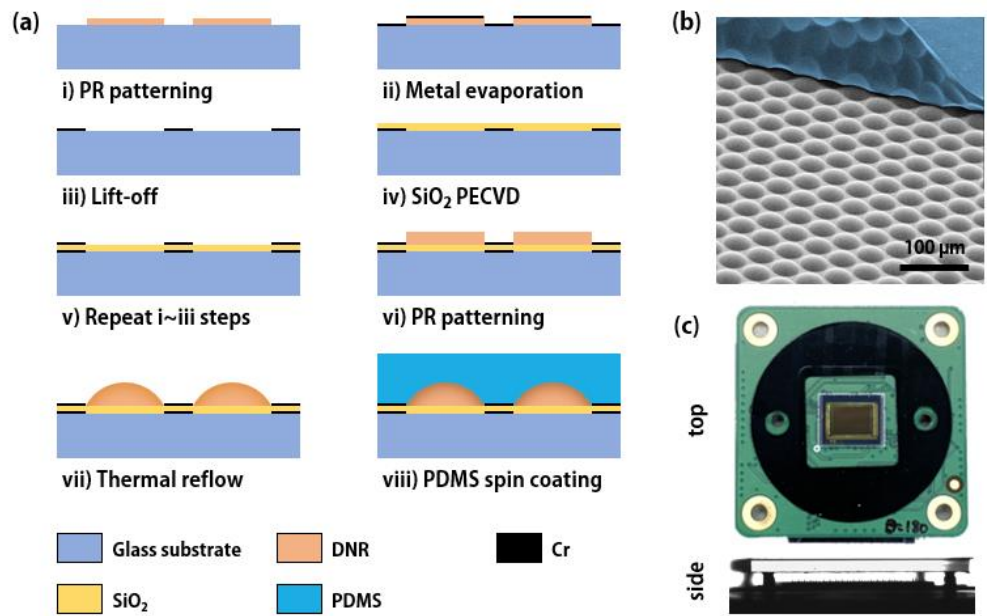

Fig. 2. (a) The microfabrication steps of siMLAs. The microlenses fabricated by thermal reflow are immersed in PDMS film by spin coating. Microlens formation by thermal reflow facilitates adding a light-absorbing layer removing optical crosstalk. (b) A scanning electron microscopy of siMLAs showing solid immersion of MLAs. The pseudo color (blue) indicates PDMS. (c) Photographs of packaged siMLAs on a CMOS image sensor using spacers.

*2.2 siMLAs characterization*

The focal length increment by solid immersion was measured by a confocal laser scanning microscope (CLSM) using a collimated light source (CPS532, Thorlabs Inc.; 532nm wavelength) (Fig. 3a). The *f*-number of MLAs is increased by immersion in PDMS by approximately 2.7 times, thereby exceeding the *f*-number limit of conventional MLAs fabricated by thermal reflow. The refractive index difference between MLAs and the immersion material dominantly determines the amount of focal length change. The refractive index of immersion material should be less than the one of microlenses, and the focal length increases more as the refractive indices of the two materials are closer. The focal length change depending on the thickness of immersion material is negligible. The transmittance of DNR film (5 μm) on a borosilicate wafer (500 μm) was measured in the visible range by a spectrometer depending on the presence of a PDMS film (10 μm) on the DNR film to compare the transmittance of MLAs and siMLAs (Fig. 3b). Refractive index buffering from the PDMS film enhances the transmittance of DNR up to 6.9% by reducing reflection from the surface. The improved transmittance of MLAs increases the contrast of the acquired images. The reflection is minimized as the refractive index of immersion material approaches the square root of the refractive index of microlenses.

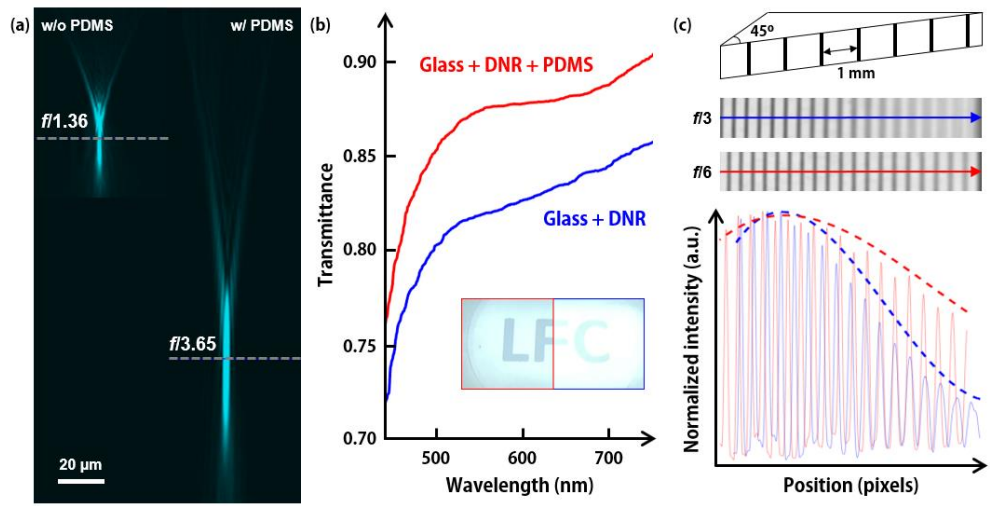

Fig. 3. Optical properties of siMLAs. (a) Cross-sectional images of laser beams focused by MLAs before and after PDMS spin coating. The focal length of microlenses is increased after solid immersion, resulting in a larger *f*-number (b) The transmittance of DNR photoresist film on a glass substrate with and without PDMS film. The optical images were taken with the printed letters on paper placed under the glass substrates. The transmittance is enhanced after coating the PDMS film on DNR by reducing reflection. (c) The reconstructed light-field of a tilted plate with a line pattern and corresponding intensity profiles captured by light-field imaging systems with different *f*-numbers. The curve fitting of peak intensities is denoted by dashed lines. The larger *f*-number brings the larger DoF in light-field imaging.

Images of a tilted plate with a line pattern were captured and reconstructed using light-field imaging systems based on siMLAs to see the DoF difference depending on the *f*-numbers (Fig. 3c). The light-field imaging system consists of a commercial standard lens (C11-2520, Basler AG; 25 mm focal length) and siMLAs mounted on an image sensor so that the systems work best at the distance of 100 mm. The tilting angle and the spacing between the line pattern on the plate are 45° and 1 mm, respectively. The captured images are reconstructed by a digital refocusing algorithm to obtain the front views of the plate. The reconstructed light-field images and their intensity profiles clearly show that the larger *f*-number is required for the larger DoF. Hence, the siMLAs with a larger *f*-number offer a larger DoF in light-field imaging compared to conventional thermally reflowed MLAs.

*2.3 3D light-field imaging application*

The siMLAs allow the design and fabrication of a compact 3D imaging system with a large DoF for biomedical applications. An LFC for 3D intraoral scanning is fabricated using a custom-made main lens, siMLAs, and an image sensor (IMX477, Sony Corp.) by rapid prototyping (Fig. 4a). The packaged intraoral scanning LFC consists of the scanner tip and the LFC module. A plane mirror is attached at the end of the scanner tip. The siMLAs of *f*/3.6 and the main lens of 28 mm focal length are packaged to make the LFC module have a working distance of 80 mm to use the apparatus conveniently imaging the inside oral cavities with the scanner tip. The modulation transfer function (MTF) was used to measure the lateral resolution and the DoF of the LFC quantitatively (Fig. 4b). Line pattern was taken from the LFC and reconstructed for MTF measurement, and the lateral resolution was calculated based on the spatial frequency at MTF50 by changing the distance between the LFC and the line pattern repeatedly. The LFC exhibits 25 μm or better lateral resolution for 10.5 mm working range. This DoF is hardly achieved by the *f*-number of conventional thermally reflowed MLAs. The 3D imaging of an actual size dental phantom has been demonstrated by the packaged LFC. The raw image was easily taken by the hand-held operation without elaborate optical stages due to enough DoF (Fig. 4c). The 4D light-field data, which contains 2D spatial and 2D angular information of light, was calculated by using the Light-Field Image Toolkit [24] from the set of microimages in the raw image. The disparity map of the captured scene was finally obtained by a cost volume-based stereo matching algorithm [25] comparing the difference in sub-aperture images (Fig. 4d). This demonstrates the possibility of compact *in vivo* 3D imaging with an LFC.

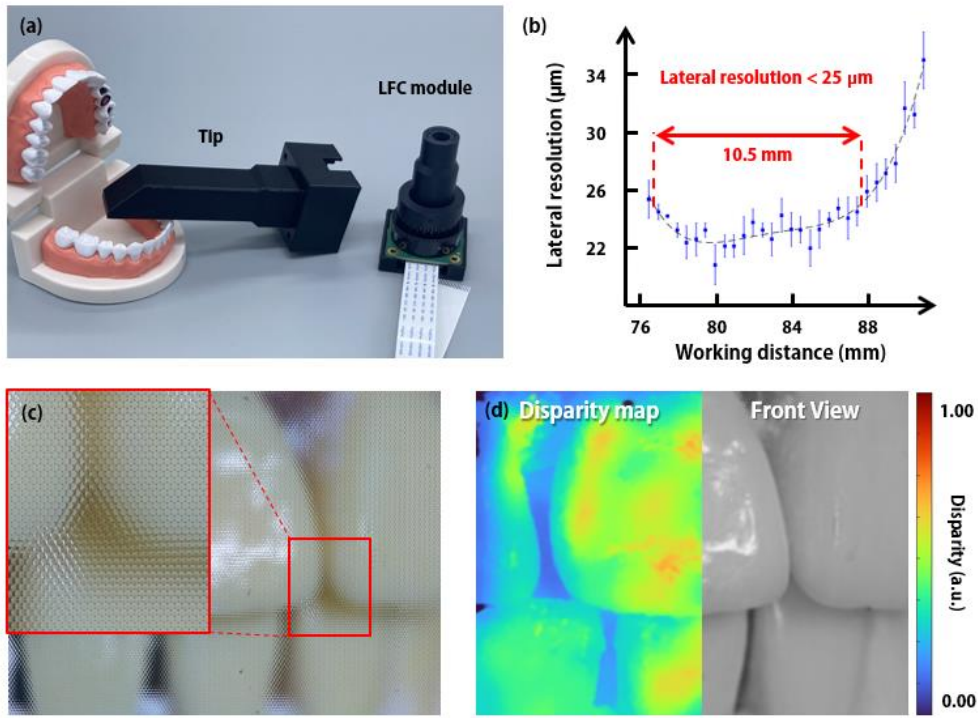

Fig. 4. Light-field imaging of an actual size dental phantom using siMLAs. (a) A photograph of the packaged LFC for 3D intraoral scanning. (b) The lateral resolution and the DoF evaluation of the LFC using MTF. (c) The raw light-field image of an actual size dental phantom captured by the LFC. (d) The light-field rendering result. The disparity map has been successfully achieved by the LFC using siMLAs.

## 3. Conclusion

In this work, the LFC based on siMLAs has been successfully demonstrated for 3D biomedical imaging with a large DoF. An additional medium between MLAs and the air increases the focal length and enhances the transmittance by utilizing the refractive index difference. The siMLAs overcome the limited *f*-number range of conventional MLAs fabricated by thermal reflow, facilitating the implementation of a light-field imaging system with a larger DoF. The advantages of thermal reflow remain after solid immersion of MLAs, such as the wafer-level fabrication and the addition of new functions for better light-field imaging. The packaged intraoral scanning LFC using siMLAs performs 3D imaging during hand-held operation through a large DoF with a practical resolution. This LFC suggests a new platform for *in vivo* imaging such as 3D endoscopy since light-field imaging enables a miniaturized 3D imaging system from its simple configuration.

**Acknowledgments.** This work was supported by the National Research Foundation of Korea (NRF) funded by the Ministry of Science and ICT (2021R1A2B5B03002428), and funded by Vatech Co. Ltd.

**Disclosures.** The authors declare no conflicts of interest.

**Data availability.** Data underlying the results presented in this paper are not publicly available at this time but may be obtained from the authors upon request.